\documentclass[a4paper]{amsproc}
\usepackage{amssymb}
\usepackage[hyphens]{url} 
\usepackage[dvips]{graphicx} 

\theoremstyle{plain}

\theoremstyle{definition}

\theoremstyle{remark}
 
 \numberwithin{equation}{section}

\title[Categorical Description Of Plant Morphogenesis]{Categorical Description Of Plant Morphogenesis}



\author[Rudskiy]{\bfseries Ivan Rudskiy} 

\address{ 
Laboratory of Scientific Projects \\ 
Saint-Petersburg\\
Russia}
\email{rudskiy@labnp.ru}

\address{
Laboratory of Embryology and Developmental Biology of Plants \\
Komarov Botanical Institute RAS \\
Saint-Petersburg \\
Russia}

\thanks{Communicated by ...} 

\begin{document}

{\begin{flushleft}\baselineskip9pt\scriptsize
\end{flushleft}}
\vspace{18mm} \setcounter{page}{1} \thispagestyle{empty}

\begin{abstract}
This article presents formalistic tool for description of structural and biochemical relations between cells in the course of development of the body of plants. This is flexible formalistic space, based on the Category theory and the Petri Net approach, which embeds and mutually supplements biological data from methodically different sources. Relation between functional and morphological ways of plant description was mathematically realized with help of the adjoint functors. It was shown that histology and proliferative activity of the formative tissues give a template for the gene interaction networks.
\end{abstract}

\maketitle

\section{Introduction}

Living plants are geometric objects with complex and diverse shape and with fine internal structure. However their development uniformly starts from the simplest unicellulular state.  It  follows the genetic program, which control the local morphogenetic processes, such as cell divisions, growth and death, via the cell-to-cell interactions organized as exchange of the chemical substances. Aim of this work is to describe the joint principles of structure, functioning and development of plants using the mathematical language.

Biology has two powerful methods of investigations : the structural-morphological approach and the molecular-genetic one. Anatomy and morphology of plants are destined to solve global problems of structural diversity and of somatic evolution. The precise and detailed classifications of plant somatic structures, such as in Esau \cite{Esau}, are among the results of these scientific directions.  More recent molecular genetic approaches are concentrated on the biochemical and logical aspects of plant development. The gene interaction networks based on the intercellular interactions are now revealed \cite{GL}. Due to the technical complexity such the works are limited only to few model plants,  such as \textit{Arabidopsis} and \textit{Zea}.

Mathematical approaches in biology allow to merge the structural globality and the molecular precision. For example, it was shown by Barlow and colleagues \cite{BLL} that the deterministic developmental processes in plant shoot meristems can be described as small programs using the grammar formalism of L-systems. 

Our formalism is based on two types of relations between cells : structural and metabolic. The spatial structure and genealogy of plant tissues was described combinatorially as graph with  associated genealogical tree. The metabolic and transport relations between cells were represented in the terms of diagrams of monoidal categories and Petri nets. 

Two almost canonically independent views to the plant existence (structural and physiological) were merged in the following way. For a given plant species the sets of all possible structural graphs and metabolic diagrams were regarded as respective categories with the functorial relations between them. Thus, the chemical reactions and traffic were considered compartmentalized according to the spatial relations between cells. Divisions and spatial relations of cells were regarded as some particular chemical reactions and control influence, respectively. As a result, the cell division activity and the cell differentiation in plant tissues were regarded as the streams of substances, enveloped into the streams of cells. Relations between these categories are regarded as the adjoint functors.

The idea of streams of substances and cells was further developed and extended to the cell-to-cell interactions. Each plant cell represents a stream of substances, namely the stream of nutrients which allow the cell to divide and to grow. Every dividing cell or a pair of the sister cells represents a stream of cells, since it produces new cells in own cell lineage. 

The streams of substances and cells interact each other by means of the exchange of nutrients and signalling molecules. The neighbour cells can influence to the division of each other and their division can influence to the neighbour cells. Thus, they perform the signal transduction like a network. 

The formalism developed in this work, describes following species-specific properties of plants :  
\begin{enumerate}
 \item Spatial arrangement of cells and their genealogy,
 \item Metabolic and signalling relations (gene interaction nets) between cells,
 \item Influence and control of the parallel morhogenetic processes in tissues.
 \end{enumerate}

The next section of this article provides a basic formalistic tools used for description of the structure and metabolism of plants  and of the developmental processes. In the section \ref{sect3} the streams of cells and substances are introduced with the respective categories. In the last section the principles of the signal transduction are shown. Several examples of the cell-to-cell signalling and influence to the cell divisions are provided.

\section{Basic formalization}

\textbf{Structural relations.} Spatial arrangement of cells in plant body is represented by graph of the spatial adjacency $G=(VG,EG)$, where the set of vertices $VG$ denotes cells and the set of edges $EG$ indicates the presence of a spatial contact or an adjacency between cells. 

\begin{figure}[htb]
\centerline{\includegraphics[width=80mm]{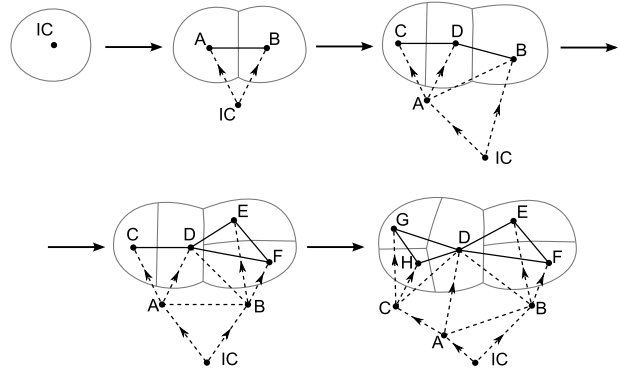}}
\caption{\small Cell divisions and transformations in the graph of spatial adjacency $G$ (solid edges) and in the genealogical tree $T$ (dashed arrows).}
\label{Fig_GT1}
\end{figure}

Developmental history of a tissue is reproduced in the form of genealogical tree, which can be reconstructed by means of the analysis of geometry of cell boundaries \cite{RH}. The genealogical tree $T$ includes living cells as pendant vertices, they coincide with the set $VG$. The other vertices represent common mother cells of the living cells and the directed edges indicate their genealogical relations. Thus, every plant organism at the particular moment of time  is structurally represented by pair $(G, T)$, see the Fig.~\ref{Fig_GT1}. 

\textbf{Metabolic relations.} We regard metabolism as a set of chemical substances with a set of chemical reactions between them. Metabolic relations between cells is a subset of substances, which are transported between cells. Chemical reactions do not preserve the history of interactions between substances. 

We use two approaches of description of metabolic relations. First is the traced monoidal categories, second is the Petri nets.

\begin{figure}[htb]
\centerline{\includegraphics[width=90mm]{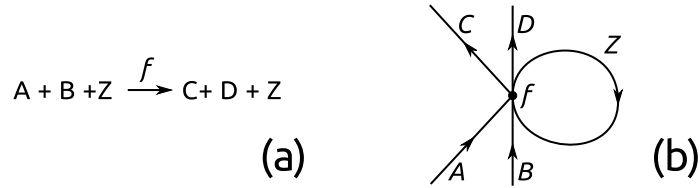}}
\caption{\small Chemical reaction (a) and its graph $C$ (b), the diagram of the traced monoidal category.}
\label{Fig_CR1}
\end{figure}
According to the first approach, metabolic relations inside and between cells are described in the form of directed graph of chemical reactions $C=(VC,EC)$. The node set $VC$ represents reactions or transformations of substances, such as substance appearance, consumption or transformation into another substance. The arrow set $EC$ represents chemical substances and the arrow direction indicates direction of substance transport (displace) in the sequence of substance transformations. Thus, all chemical reactions in every plant organism at the particular moment of time are represented as possibly non-connected graph $C$ (Fig.~\ref{Fig_CR1}).

\begin{figure}[htb]
\centerline{\includegraphics[width=99mm]{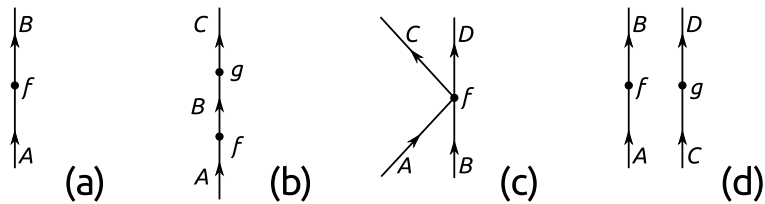}}
\caption{\small Diagram of the monoidal category. (a) -- objects $ A, B$ and morphism $f:A\to B$. (b) -- composition of two morphism $fg:A\to C$. (c) -- morphism of tensor products $f: A\otimes B\to C\otimes D$. (d) -- tensor product of two morphisms $f\otimes g$.}
\label{Fig_TMC1}
\end{figure}

Graph $C$ is canonically considered a diagram of the traced monoidal category in the following way \cite{JSV,S}. Objects correspond to arrows, morphisms to vertices (Fig.~\ref{Fig_TMC1}a). The identity morphism is an arrow with the vertex omitted. The composition of two morphisms is a concatenation of three arrows (Fig.~\ref{Fig_TMC1}b). Tensor product and other properties of the monoidal categories are defined in a standard way (Fig.~\ref{Fig_TMC1}c,d).

A trace for a monoidal category $C$ is a family of functions  \cite{JSV}:
\begin{equation*}
\mathrm{Tr}^{Z}_{A\otimes B,C\otimes D} : \mathrm{Hom}_{c}(A\otimes B\otimes Z, C\otimes D\otimes Z) \longrightarrow \mathrm{Hom}_{c}(A\otimes B, C\otimes D)
\end{equation*}
The trace is denoted diagrammatically as a loop, where it corresponds to the catalytic function of the substance ($Z$), which is necessary for the chemical reaction, but is kept unchanged (Fig.~\ref{Fig_CR1}b). The trace function gives several axioms in addition to the common ones, such as vanishing, strength, sliding (see \cite{S} for more details).

A diagram of the traced monoidal category represents a system of all the chemical reactions possible for a given plant, i.e. it shows statics. The dynamics of a system, namely the presence or absence of reagents in the reaction media and their traffic is reproduced by means of the Petri net approach developed for description of various dynamical systems \cite{P,M}. 

\begin{figure}[htb]
\centerline{\includegraphics[width=80mm]{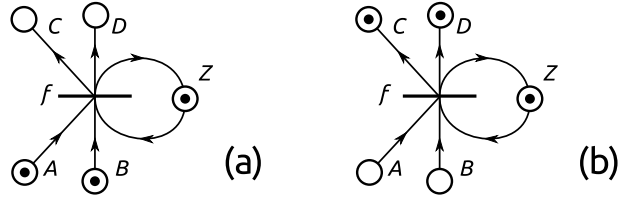}}
\caption{\small Petri net for the chemical reaction $A+B+Z\to C+D+Z$. (a) -- initial state, transition $f$ is enabled. (b) -- state after execution of the transition $f$.}
\label{Fig_PN1}
\end{figure}

A Petri net is an oriented graph with two types of vertices ``places'' and ``transitions'', and arrows. The set of places corresponds to the set of properties of a dynamical system, in our case it is the chemical substances. The set of transitions corresponds to the set of transformations, actions of a system. Here,  it is a set of chemical reactions. The set of arrows denotes the control flows. Every arrow connects a place with a transition or vise verse. In our formalism, this is a correspondence between substances and chemical reactions (Fig.~\ref{Fig_PN1}). 

A graph of the Petri net has a marking function. The vertices from the subset ```places'' has a mark called ``token''. Presence or absence of tokens on a place denotes the presence or the absence of the respective property in system. Vertices from the subset ``transitions'' serve as rewriting functions for the marking. A transition takes tokens from the sources of its incoming arrows and adds tokens to the place at the end of every its outgoing arrows. The availability of the tokens in every the place of the incoming arrows is a condition for the transition to be executed (Fig.~\ref{Fig_CR1} a,b). Such a transition is called ``enabled''. 

\textbf{Correspondence between monoidal categories and Petri nets.} We use the following correspondence between our two formalistic approaches for metabolic relations. The sets of objects and morphisms in a monoidal category coincides, respectively,  with the set of places and transitions in a  Petri net. Tensor product on $n$ objects or $m$ morphisms in a monoidal category is a set of $n$ places or, respectively,  $m$ transitions in a Petri net. 

\begin{figure}[htb]
\centerline{\includegraphics[width=99mm]{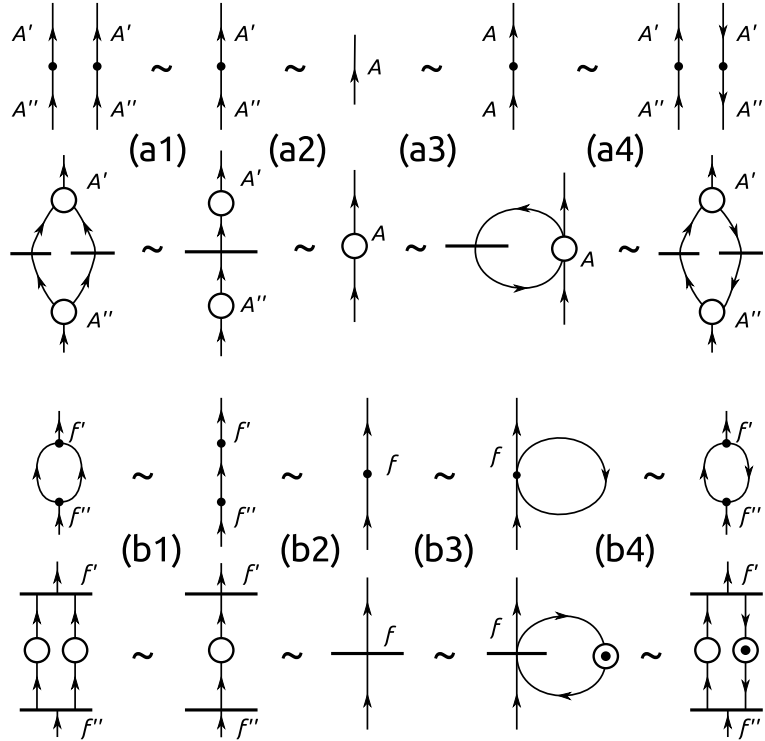}}
\caption{\small Transformations of the diagrams of monoidal categories (above) and of the Petri nets (below), which preserve properties of a system. (a) -- for the vertices (transitions). (b) -- for the edges (places).}
\label{Fig_PN2}
\end{figure}

Diagrams of the monoidal categories and the Petri nets admit structural transformations which do not change their structural and dynamical properties, such as connectedness, safeness, boundedness and liveness. These transformations are known to be isomorphisms for the graphs and refinements for the Petri nets \cite{SM}. These transformations include fusions of the parallel vertices and series of the vertices (transitions) and the same for the edges (places), elimination of the self-loops and identities. All possible transformations are shown for the diagrams and for the Petri nets (Fig.~\ref{Fig_PN2}).

\textbf{Structural and metabolic transformations.} There are four general morphogenetic processes responsible for the structural and metabolic transformations of plant tissues : 
\begin{enumerate}
 \item Cell division (mitosis), augmentation of the cell number;
 \item Cell growth, expansion or stretching;
 \item Cell differentiation, metabolic transformation of a cell;
 \item Cell death, the process of cell elimination.
 \end{enumerate}

Formally, cell division is a substitution of a vertex $v$ in the graph of adjacency $G$ by the pair $v'$, $v''$ and by the edge $v'v''$ between them. Edges of the vertex $v$ are distributed or split between $v'$ and $v''$. The respective transformation is happening with the genealogical tree $T$. See the Fig.~\ref{Fig_GT1}. Cell death is a process inverse to the cell division. Cell growth and differentiation do not influence to the pair $(G,T)$.

\begin{figure}[htb]
\centerline{\includegraphics[width=80mm]{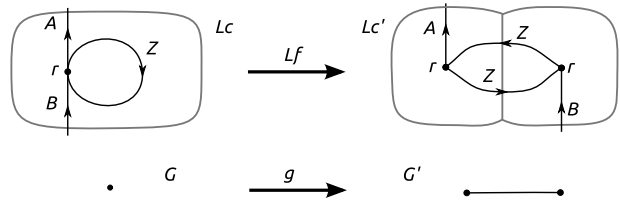}}
\caption{\small Transformation of metabolic and spatial relations in the course of cell division.}
\label{Fig_TMC2}
\end{figure}

Each morphogenetic process is connected with the transformation of the transport relations between cells. In case of the cell division, it is a redistribution or a reproduction of the distinct  chemical processes or of the transport relations. The metabolic transformations do not alter the set of chemical reactions and the transport relations of a whole plant. Formally, it means that the diagram transformations are the isomorphisms of graphs, which are based on the splits and the fuses  of the labelled vertices and the edges (Fig.~\ref{Fig_TMC2}). 

The boundary of each cell represents a set of transformations of substances associated with the intercellular transportation (Fig.~\ref{Fig_TMC2}). There are input and output transformations, some of them connect a cell to the external environment, some of them do to the neighbour cells. It is obvious that between any two cells in plant body there is a traffic relation appeared in the result of the cell divisions happened with their common mother cells. Thus, the plant body is a traffic network between the integrated cellular transport compartments or domains. More details about this can be found in Gamaley \cite{G}.

\begin{figure}[htb]
\centerline{\includegraphics[width=80mm]{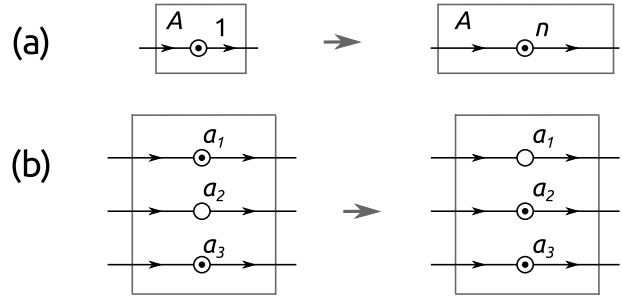}}
\caption{\small Representation of cell growth (a) and differentiation (b).}
\label{Fig_SG1}
\end{figure}

Cell growth and differentiation are dynamical properties of the metabolic relations. Cell growth means that some substances are collected in the cell (Fig.~\ref{Fig_SG1}a). Cell differentiation is a change of the metabolic profile. A cell can cease consumption and emission of one substance and can start exchange of the another one (Fig.~\ref{Fig_SG1}b). In the real plant tissues different morphogenetic processes accompany each other. 

\section{Concordance of developmental transformations in the tissue structure and transport}\label{sect3}

\textbf{Streams of cells and substances.} In plant bodies there are two main types of tissues : the formative tissues, with dividing cells, and the somatic, differentiated tissues with physiologically active cells. Formative tissues produce new cells for the somatic tissues. Somatic tissues support the formative tissues with nutrients. They have limited period of life span and need to be regularly renovated.  

Structure and dynamics of formative and somatic tissues determine appearance of a whole plant. Formative tissues prevails in embryo and in developing organs. Renovation of tissues and organs is necessary for all the perennial plants to keep the transport and assimilative functions of their bodies. Senescence and programmed death of cells are regular processes in somatic tissues. More detailed descriptions of the regularities in embryogenesis and morphogenesis of plants can be found in \cite{Esau} and \cite{Sh}.

We consider two streams in plant body. The stream of cells, which is started in the formative tissues and is directed to the somatic tissues. And the stream of substances, which goes from the somatic tissues and is directed to the formative tissues.

\begin{figure}[htb]
\centerline{\includegraphics[width=99mm]{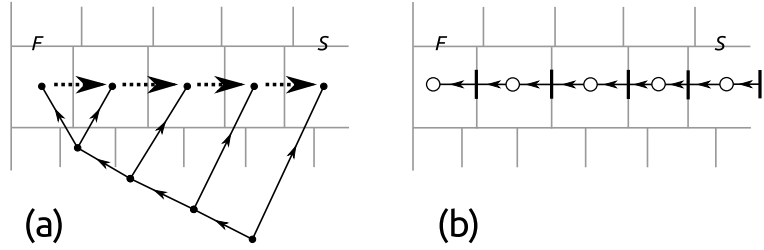}}
\caption{\small Streams of cells (a) and substances (b) between formative (\textit{F}) and somatic (\textit{S}) tissues on the example of a single cell lineage.}
\label{Fig_SCS1}
\end{figure}

These two streams support each other in a natural way.  The somatic tissues are supplemented by newly differentiated cells from the formative tissues (Fig.~\ref{Fig_SCS1}a). These new cells start their functioning and produce nutritive substances for the dividing cells (Fig.~\ref{Fig_SCS1}b).

It is possible to regard the stream of cells in the terms of the stream of substances. The intercellular traffic is a stream of substances, which are used for the growth and the signalling. The stream of cells is a stream of substances compartmentalized inside the limits of cells. The destruction of cells only partially releases substances back to the transport network. Thus, the cellular body of plants is form of accumulation of substances, as a balance between growth and degradation. 

\textbf{Categorical description of spatial and transport relations in tissues.} A category of spatial and genealogical relations $\textbf{S}$ is a category of graphs with : 
 \begin{enumerate}
  \item pairs $s=(G,T)$, for a plant body or its fragment at every moment of time, as objects,
  \item developmental transformations occurred between two successive moments $g: (G,T) \longrightarrow  (G',T')$, as morphisms,
  \item trivial transformation $\text{Id}(s): (G,T) \longrightarrow  (G,T)$, as identity,
  \item extension of the developmental transformation for successive moments  $g_{1}g_{2}: (G_{0},T_{0}) \longrightarrow  (G_{1},T_{1}) \longrightarrow (G_{2},T_{2}) = (G_{0},T_{0}) \longrightarrow (G_{2},T_{2})$, as composition of morphisms. 
\end{enumerate}

There is partial ordering between some objects in $\textbf{S}$. Two objects are related $s_{1}\geqslant s_{2}$, if genealogical tree $T_{2}$ is a subtree of tree $T_{1}$ and graph $G_{2}$ represents the arrangement of the mother cells of the cells described in graph $G_{1}$. This relation defines classes of equivalences (classes of parents) of the cell arrangements in $\textbf{S}$.

A category of metabolic relations $\textbf{C}$ consists of :
\begin{enumerate}
  \item diagrams $c$ of the monoidal category or the Petri net assigned to the metabolic relations in a plant at a given moment of time, as objects,
  \item possible isomorphic transformations of diagrams $f: c \longrightarrow  c'$, which correspond to the transformations of metabolic relations, as morphisms,
  \item trivial diagram transformation $\text{Id}(c): c \longrightarrow  c$, as identity,
  \item extension of the developmental transformation for successive moments  $f_{1}f_{2}: c_{0} \longrightarrow  c_{1} \longrightarrow c_{2} = c_{0} \longrightarrow c_{2}$, as composition of morphisms. 
\end{enumerate}

There is partial ordering between some objects in $\textbf{C}$. Two objects are related $c_{1}\geqslant c_{2}$, if $c_{2}$ represent a pathway in the diagram $c_{1}$. This relation defines classes of equivalences (classes of transportation) of the metabolic relations in $\textbf{C}$.

\textbf{Relations between streams of cells and substances.} There are two equal points of view to the morphogenetic processes in plants : structural and physiological. From the structural point of view a cell represents interacting streams of substances, a location of their metabolic exchange. From the other point of view, tissues are the streams of cells which service all biological needs of plants. In order to describe equivalence of theses views we are going to construct the functors between categories $\textbf{S}$ and $\textbf{C}$ :
\begin{equation*}
 L: \textbf{C} \longrightarrow  \textbf{S}\; \qquad  \text{and} \qquad R: \textbf{S} \longrightarrow  \textbf{C}
\end{equation*}

A diagram $c \in \textbf{C}$ defines a class $Lc$ of transport compartmentalizations in $\textbf{S}$ as follows. The object $Lc$ is graph of spatial adjacency, where each the vertex $v \in VLc$ is a subdiagram of $c$ and each edge $e=vv' \in ELc$ is set of transformations of substances associated to the boundary between $v$ and $v'$ (Fig.~\ref{Fig_TMC2}). Class $Lc$ is defined as a set of all possible compartmentalizations with the same scheme of metabolic relations. The set of metabolic ends in the diagram $c$ corresponds to the set of the sister cells in each compartmentalization in $Lc$.

The object $Lf: Lc \to Lc'$ is a class of possible developmental transformations in the cellular architecture which admit the respective isomorphism of transport diagrams $f:c \to c'$.

A cellular architecture $s \in \textbf{S}$ defines a class $Rs$ of metabolic diagrams in $\textbf{C}$ in the following way. The object $Rs$ is a transport (metabolic) diagram, where each the cell $v \in VG$ represents a subset of substances and their transformations $\{f_i\}$ inside this cell $Rv=f_{1}\otimes f_{2} \otimes ...$, and each the edge $vv'\in EG$ is a set of chemical reactions $\{f'_j\}$ on the boundary of two cells $v$ and $v'$ $Rvv'=f'_{1}\otimes f'_{2} \otimes ...$. For the given cellular architecture $G$ and developmental history $T$, the class $Rs$ is defined as a set of all possible metabolic networks realizable by means of transformations, which preserve the substances and their pathways in a system (see Fig.~\ref{Fig_PN2} a1, a3). 

The morphism $Rg: Rs\to Rs'$ is a transformation of the metabolic relations in the concordance with the new graph of spatial adjacency $G'$ and the genealogical tree $T'$. The class $Rg$ is defined as a set of all possible transformations of the metabolic network $Rs$, which change the number of substances (see Fig.~\ref{Fig_PN2} b1--b4, a2, a4). Thus, each plant cell $v \in VG$ is regarded under the functor $R$ as complex chemical substance $Rv\in RVG$ and each cell division is regarded as complex chemical reaction $Rvv'\in REG$ associated to the boundary between two daughter cells $v$ and $v'$ (chemical reaction $x_{i}+a_{j}\to b_{j'}$ in the Fig. \ref{Fig_Rs2b}).

Two functors $L$ and $R$ are left and right adjoint functors, respectively \cite{ML}. There are the following relations and equivalence :
\begin{equation}\label{eq:1}
 Lc \geqslant s \; ; \qquad c \leqslant Rs \; ; \qquad  \text{Hom}_{\textbf{S}} (Lc, s) \approx\; \text{Hom}_{\textbf{C}} (c, Rs).
\end{equation}

The relations between two categories $\textbf{S}$ and $\textbf{C}$ mean that a given stream of cells $s=(G,T)$, with the respective genealogical tree and the spatial arrangement of cells, contains unique and species-specific realization of the diagram of metabolic relations $c$ in the form of the stream of cells $Lc$. And vice verse, the metabolic relations represented in the diagram $c$ are uniquely and species-specifically realized in the class of streams of substances $Rs$. In the other words, equality \eqref{eq:1} means species-specific correspondence between metabolic diagrams and cellular architecture in the languages of cells and transport system.

\section{Interaction of developmental processes and their control}

\textbf{Signal transduction network.} Cell communication or signalling is necessary for proper development of plant tissues and organs. This is realized by means of exchange of substances between cells. Sugars and other metabolites are mainly used for the long-distance (global) signalling, while proteins and hormones have more important and precise action in the local scale \cite{G}.  

Local intercellular communication involves signalling between organelles and common transport systems of neighbour cells, such as the reticulum and the cytoplasm. There is a group of specific proteins, factors of transcription, which can be transported between cells. These molecules positively or negatively control gene activity in a cell and thus, they influence to the metabolism and mitotic activity.  This class of chemical interactions is called gene regulatory networks \cite{GL}.

There are direct relations between structure and chemical processes, between morphogenetic processes and transport of substances in plant tissues. According to the formalism of the adjoint functors $L$ and $R$, the streams of substances  and cells, with their respective representatives as classes $Lc$ and $Rs$, are intrinsically the same things characterized from the different points of view. Below we continue to describe developmental processes in the terms of the functor $R$, namely,  we regard cells and cell divisions as special chemical substances and chemical processes, respectively, and streams of cells as compartmentalized streams of substances. 

\begin{figure}[htb]
\centerline{\includegraphics[width=85mm]{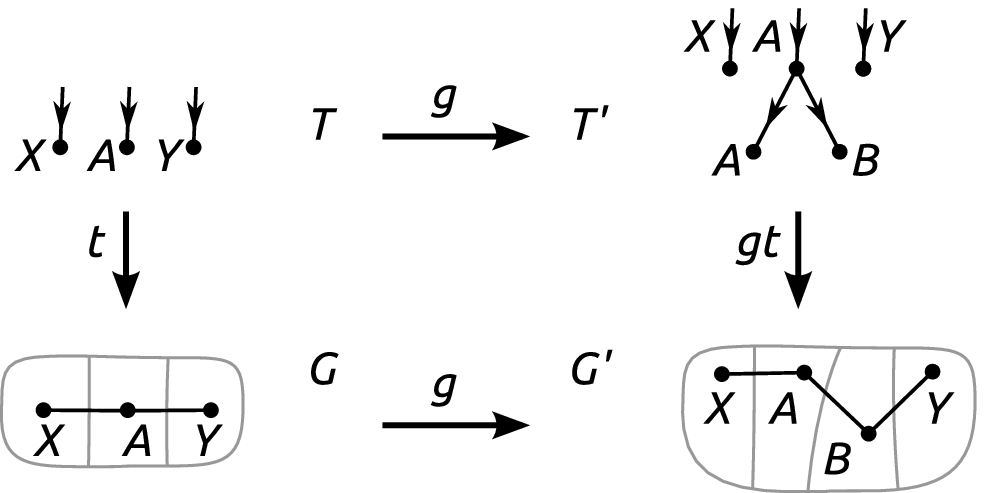}}
\caption{\small Morphism $g:s\to s'$ in the category $\textbf{S}$ and associated map map $t:T\to G$.}
\label{Fig_Rs2}
\end{figure}

The developmental history of a tissue is a morphism $g:s\to s'$ and the genealogical tree $T$ is a path of the stream of cells. This defines a surjective map $t:T\to G$, such that $T$ is projected to the graph of adjacency $G$. Every edge of $G$ is an image of the respective edge of the tree $T$ appeared after the cell division(s) $g$ (edge $AB$ in $T$ and $G$, for example, in the Fig. \ref{Fig_Rs2}). 

\begin{figure}[htb]
\centerline{\includegraphics[width=90mm]{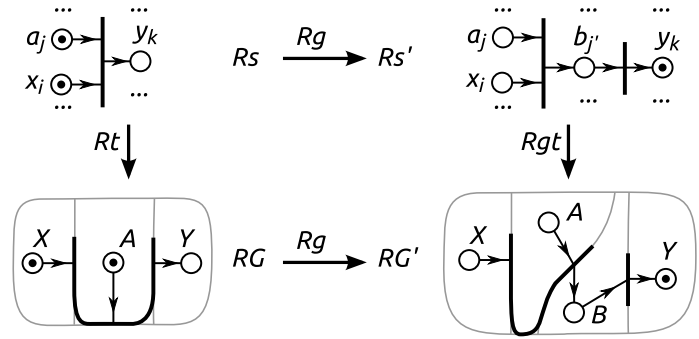}}
\caption{\small Morphism $Rg:Rs\to Rs'$ and function $Rt:Rs\to RG$ in the category $\textbf{C}$. Cell division accompanies transport of substances.}
\label{Fig_Rs2b}
\end{figure}

The functor $R$ preserves these  maps as follows. In the language of chemical processes, the adjacency graph $G$ represents a class of diagrams $Rs$ and the genealogical tree $T$ defines classes of paths of tokens in $Rs$, which are involved into the stream of cells. Morphism $Rg:Rs\to Rs'$ describes cell divisions in the terms of transformations of transport diagrams.  The map $Rt:Rs\to RG$ describes detailed distribution of substances between cells, which was set up after each developmental transformation $g$ (Fig. \ref{Fig_Rs2b}), according to the equality \eqref{eq:1}. Thus, the category $C$ acquires the property of the process history.

The object $Rs$ is a network of chemical relations inside and between cells. It contains any admissible molecular genetic regulatory network and any transport network peculiar for the given plant species at the particular moment of time. The object $RG$ is a compartmentalization of $Rs$, such that every place $Rv$ contains non-empty places of $Rs$ (in the form of the Petri net) and the subset of transitions $REG$ includes reactions associated to the cell boundaries. 

Let us regard the dynamics in the diagrams $Rs$ and $RG$. The signal in $Rs$ and in $RG$ is some amount of chemical substances, a token. In $Rs$ the signal is transformed by chemical reactions, thus providing the dynamics. The dynamics in $RG$ represents morphogenetic processes based on the cell divisions and the associated intercellular transport. A cell is capable to directly participate in own cell division and create two daughter cells, or it can influence to the division of its neighbour cells by sending signal substances (Fig.~\ref{Fig_Rs2b}).

\begin{figure}[htb]
\centerline{\includegraphics[width=80mm]{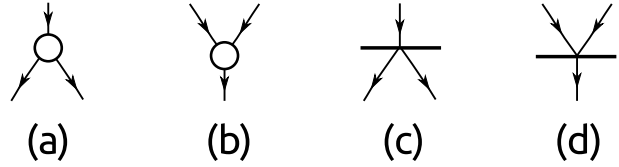}}
\caption{\small Process branching and fusion on places (a, b) and on transitions (c, d).}
\label{Fig_PN3}
\end{figure}

There are four types of the signal (process) interaction in Petri nets:
\begin{enumerate}
  \item Choice or conflict, branching on a place. Not determinate selection of a way for the process continue (Fig.~\ref{Fig_PN3}a).
  \item Not determinate merge, fusion on a place. Not determinate prefix for a process (Fig.~\ref{Fig_PN3}b).
  \item Concurrent or parallel processes, branching on a transition. Parallel processes are initiated  by transition execution (Fig.~\ref{Fig_PN3}c).
  \item Synchronization of processes, fusion on a transition. The process will be continued, only if the tokens are available in all the merging branches (Fig.~\ref{Fig_PN3}d)
\end{enumerate}

Basic properties of the different signal interactions will be demonstrated on the following examples.

\begin{figure}[htb]
\centerline{\includegraphics[width=99mm]{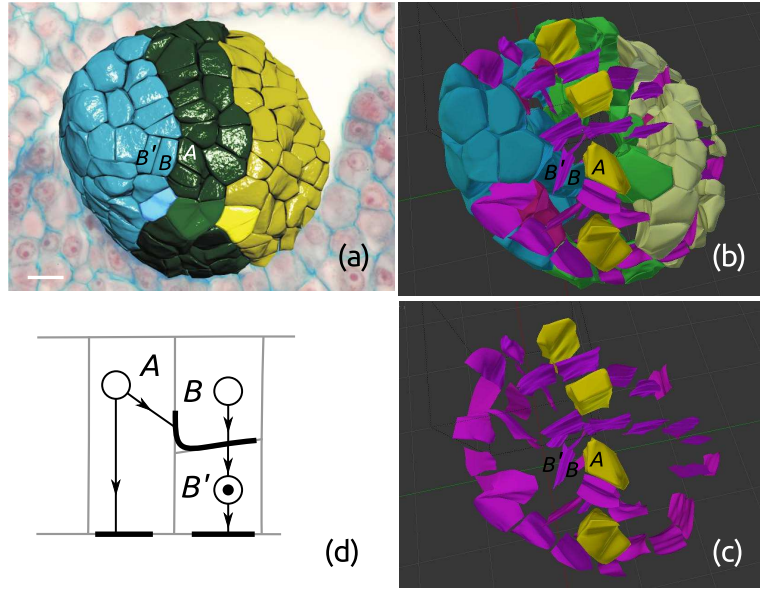}}
\caption{\small Dynamics of cell divisions in the shoot apex in \textit{Calla palustris}. (a) -- reconstruction of the cellular architecture. (b) -- tissue with the dividing cells of type B shown as planes of divisions (violet), non-dividing cells of type A are shown yellow. (c) -- cells of types A, B (planes of divisions). (d) -- diagram of the signalling.}
\label{Fig_Calla}
\end{figure}

\textbf{Dynamical state of the traffic in the shoot apex in \textit{Calla palustris}}. We have analyzed the three-dimensional structure of the vegetative shoot apex in \textit{Calla palustris} L., Araceae (\cite{RH}; Fig.~\ref{Fig_Calla}a). There are various signalling relations between cells in the apex. Most remarkable are the four non-dividing cells (type  A), each of them is surrounded by up to 8 pairs of the dividing cells (type B, see the Fig.~\ref{Fig_Calla}). These stagnant cells with their local environments composed of the active cells create the magnificent honey-comb-like proper structure (Fig.~\ref{Fig_Calla}c). 

Divisions of the cells of type A are suppressed. These cells control divisions of the neighbour cells of type B by sending them a signal, which is synchronized with their division. The cells of type A demonstrate not determinate choice via influence to the division of the adjacent cells of type B instead of own division (Fig.~\ref{Fig_Calla}d). This complex of cells with such a dynamics of divisions is responsible for the regular successive formation of the new leaves on the shoot apex in \textit{Calla palustris}.

\begin{figure}[htb]
\centerline{\includegraphics[width=99mm]{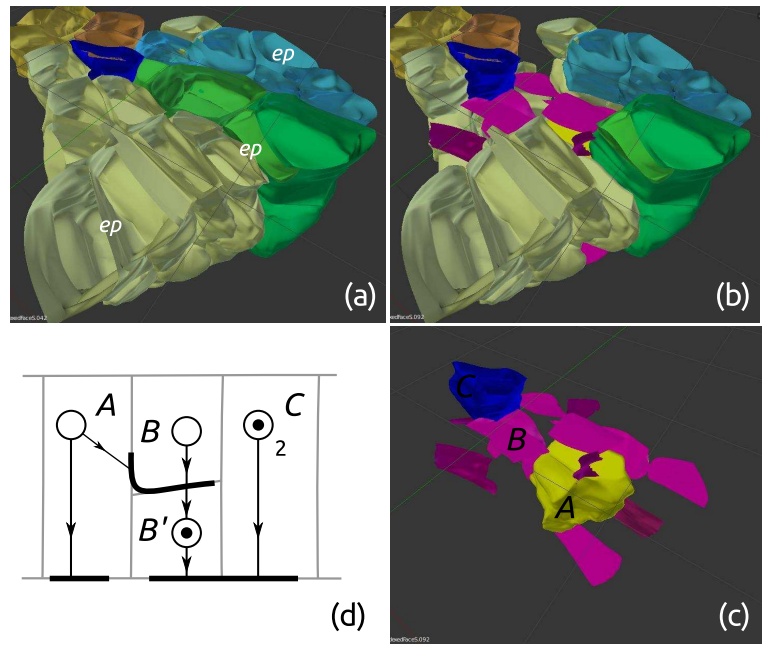}}
\caption{\small Dynamics of cell divisions in the ovule primordium in \textit{Dioscorea caucasica}. (a) -- reconstruction of the cellular architecture, \textit{ep} -- epiderm. (b) -- tissue with the dividing cells of type B shown as planes of divisions (violet), non-dividing cells of type A are shown yellow. (c) -- cells of types A, B (planes of divisions) and C (dark blue). (d) -- diagram of the signalling.}
\label{Fig_Dioscorea1}
\end{figure}

\textbf{Dynamical state of the traffic in the ovule primordium in \textit{Dioscorea caucasica}}. The analogous work was done with a primordium of the ovule in \textit{Dioscorea caucasica} Lipsky (Dioscoreaceae) (\cite{TRS}; Fig.~\ref{Fig_Dioscorea1}a). The ovule is a generative structure destined for the sexual reproduction and for the creation of the embryo sac and further the seed. 

The archaesporic  cell is mother cell for the embryo sac, it fulfils the role of the cell of type A with suppressed cell divisions. This cell activates divisions in the surrounding cells of type B. In this system there is a cell of type C, whose divisions are synchronized with each second or third division of the neighbour cells of type B. The volume of this cell is respectively augmented in two or three times (Fig.~\ref{Fig_Dioscorea1}). All this cells are important histogenetic initials, which are responsible for creating of the sophisticated reproductive organ, an ovule. 

Diagrams of the signalling in both examples can serve as templates for the revelation of the species-specific gene interaction networks.

\section*{Acknowledgements}

I express my profound acknowledgements to my former colleagues from Institut des Hautes \'{E}tudes Scientifiques, Bures-sur-Yvette, France : Morozova N., Gromov M., Bourguignon J.P., Kontcevich M., who helped and supported me much in my work for the long period of time. 

I would like to thank organizers of the XIX Geometric Seminar in Zlatibor,  2016, for their interest to my work and hospitality. My special thanks to M. \DJ{}ori\'{c} and Z. Raki\'{c}.


\bibliographystyle{amsplain}

\end{document}